\documentstyle[12pt]{article}

\title{Translations between Quaternion and Complex Quantum Mechanics}
\author{by\\ {\bf S. De Leo} and  {\bf P. Rotelli}\\ Dipartimento
di Fisica, Universit\`a  di Lecce\\ and INFN-Sezione di Lecce}
\date{14 December 1993}

\begin{document}

\maketitle

\begin{abstract}
While in general there is no one-to-one correspondence between complex and
quaternion quantum mechanics (QQM), there exists at least one version of QQM
in which a {\em partial} set of {\em translations} may be made. We define
these translations
and use the rules to obtain rapid quaternion counterparts (some of which
are new) of standard quantum mechanical results.
\end{abstract}

\vspace*{20mm}

Bitnet:DELEOS@LECCE.INFN.IT

\pagebreak

\hspace*{5mm} In this letter we wish to exhibit explicitly a set of rules
for passing back and forth between standard (complex) quantum mechanics
and the quaternion version of ref.\cite{rot1}. This will not be possible in
all situations, so this ``translation'' is only partial, consistent with
the fact that the quaternion version (QQM) provides additional physical
predictions. In a pure translation nothing can be predicted which is not
already in the original theory, although some assumptions may
appear more or less ``natural'', some calculations may be more or less
rapid and some (new) results may appear in the translated version for the
first time.

In the work of ref.\cite{rot1} a quaternion version of the Dirac equation was
derived (see also ref.\cite{mor1}) in the form
\begin{equation}
\gamma^{\mu} \partial_{\mu} \psi i = m \psi
\end{equation}
where the $\gamma^{\mu}$ are two by two quaternion matrices satisfying the
Dirac condition
\begin{equation}
\{ \gamma^{\mu}, \gamma^{\nu} \} = 2g^{\mu \nu}
\end{equation}
and the adjoint matrix satisfies,
\begin{equation}
(\gamma^{\mu})^{+} = \gamma^0 \gamma^{\mu} \gamma^0 \; \: .
\end{equation}
In this formalism the momentum operator must be defined as
\begin{equation} \label{a}
p^{\mu} = \partial^{\mu} \vert i
\end{equation}
where a bared operator $A\vert b$ acts as follows upon the
quaternion spinor (column matrix) $\psi$
\begin{equation}
(A\vert b) \psi \: \equiv \: A \psi b \; \: .
\end{equation}
We anticipate that to date the only $b$ term that has appeared in this
formalism is $i$ (except of course for the trivial identity).

The three-momentum part of the operator in eq.(\ref{a}) is hermitian only
if for the scalar product one adopts the complex scalar product\cite{hor}
(CSP). This choice is generally accepted in order to  be able to define
the quaternion tensor product and one of the results of this paper will be to
derive the rules for performing quaternion tensor products directly
from the rules for the tensor products of standard quantum mechanics
after ``translation''.

The use of quaternions in quantum mechanics may be suggested by analogy
between the imaginary units $(i, j, k)$ of a general quaternion $q$
\begin{equation}
q = q_{0} + q_{1} i + q_{2} j + q_{3} k
\end{equation}
\begin{center}
$(q_{m}\in {\cal }R \hspace*{5mm} m=0, \ldots, 3)$
\end{center}
with
\begin{equation}
i^{2} = j^{2} = k^{2} = -1 \: ; \hspace*{10mm} i j k = -1 \: .
\end{equation}
and the Pauli sigma matrices $\sigma_{m}$. More correctly given the
hermitian properties of $\sigma_{m}$, the analogy should be made with
$-i \sigma_{m}$ or a similar set. Normally the difficulty of using
quaternions in quantum mechanics arises in the question of what represents
the ``$i$'' in the dynamical equations of fermionic particles or indeed in
the momentum operator or in the Heisenberg uncertainty relation etc. In the
version that we have studied in the past, the first steps of which have been
outlined above, this task is performed by $1\vert i$. Obviously the identity
of the right acting $i$ with that from the left is not without physical
consequences.
For one thing it breaks the symmetry amongst the imaginary quaternion
units and justifies the use of a preferred complex plain for the scalar
product. We note in passing that our approach is different from that of
Morita (see ref.\cite{mor2}) who uses complex
quaternions (or biquaternions) which contain an
additional {\em commuting} imaginary ${\cal I} = \sqrt{-1}$.

Returning to our Dirac equation we note that $\psi$ is a quaternion two
component spinor, which yields {\em four} plain wave solutions which are
complex orthogonal (n.b. that $1$ and $j$ are complex orthogonal numbers).
To complete this introduction we also recall that with this formalism, new
physics appears in the bosonic sector if, as is natural, we adopt the
standard wave equations Klein-Gordon, Maxwell etc. In these cases, we find a
doubling of the standard complex solutions and hence the appearance of
``anomalous'' solutions\cite{del1,rot2}.

Normally the distinction between operators and states is manifest as in
standard complex quantum mechanics. Only with simple (one-dimensional)
quaternions is there any need to specify explicitly the difference. We
begin by recalling the so called ``symplectic'' complex representation of a
quaternion (state) $q$ \cite{hor}
\begin{equation} \label{b}
q = a + j \tilde{a} \hspace*{15mm} a, \tilde{a} \in {\cal C}(1,i)
\end{equation}
by the complex column matrix
\begin{eqnarray} \label{c}
q \: \leftrightarrow \: \left( \begin{array}{c} a \\
\tilde{a}  \end{array} \right) \; \: .
\end{eqnarray}
We now identify the operator representations of $i, j$ and $k$ consistent
with the above identification:
\begin{eqnarray} \label{d}
\begin{array}{ccccr}
i & \leftrightarrow & \left( \begin{array}{cc} i & 0 \\
0 & -i\end{array} \right) & = & i\sigma_{3} \\ \\
j &  \leftrightarrow &  \left( \begin{array}{cc} 0 & -1 \\
1 & 0\end{array} \right) & = & -i\sigma_{2} \\ \\
k &  \leftrightarrow &  \left( \begin{array}{cc} 0 & -i \\
-i & 0\end{array} \right) & = & -i\sigma_{1} \\ \\
1 & \leftrightarrow & \left( \begin{array}{cc} 1 & 0 \\
0 & 1\end{array} \right) & & \end{array}
\end{eqnarray}
e.g. it is readily checked that the ``state''
\[ jq \: \leftrightarrow \: \left( \begin{array}{c} -\tilde{a} \\
a \end{array} \right) \]
however it is calculated.

The translation in eq.(\ref{d}) (or equivalent) has been known since the
discovery of quaternions. It permits any quaternion number or matrix (by
the obvious generalization) to be translated into a complex matrix, but not
necessarily vice-versa. Eight real numbers are necessary to define the most
general $2 \times 2$ complex matrix but only four are needed to define the
most general quaternion. Infact since every (non-zero) quaternion has an
inverse, only a sub-class of invertible $2 \times 2$ complex matrices are
identifiable with quaternions.

To complete the translation we therefore need four additional degrees of
freedom, these can be identified with
\begin{eqnarray} \label{e}
\begin{array}{ccc}
1\vert i & \leftrightarrow & \left( \begin{array}{cc} i & 0 \\
0 & i\end{array} \right)\\ \\
i\vert i & \leftrightarrow & \left( \begin{array}{cc} -1 & 0 \\
0 & 1\end{array} \right)\\ \\
j\vert i & \leftrightarrow & \left( \begin{array}{cc} 0 & -i \\
i & 0\end{array} \right)\\ \\
k\vert i & \leftrightarrow & \left( \begin{array}{cc} 0 & 1 \\
1 & 0\end{array} \right)
\end{array}
\end{eqnarray}

It is readily seen that the definitions (\ref{b}), (\ref{c}), (\ref{d}) and
(\ref{e}) are all consistent with each other, e.g.
\begin{eqnarray*}
j \times 1\vert i \: = \: j\vert i & \Leftrightarrow & \left( \begin{array}{cc}
 0 & -1 \\ 1 & 0\end{array} \right)  \left( \begin{array}{cc} i & 0 \\
0 & i\end{array} \right) \: = \: \left( \begin{array}{cc} 0 & -i\\
i & 0\end{array} \right)
\end{eqnarray*}
and
\begin{eqnarray*}
k\vert i \times j\vert i \: = \: kj\vert i^{2} \: = \: -kj
\: = \: i & \Leftrightarrow & \left( \begin{array}{cc} 0 & 1 \\
1 & 0\end{array} \right)  \left( \begin{array}{cc} 0 & -i \\
i & 0\end{array} \right) \: = \: \left( \begin{array}{cc} i & 0\\
0 & -i\end{array} \right) \: \: .
\end{eqnarray*}
With these rules we can translate any quaternion matrix operator into an
equivalent {\em even dimensional} complex matrix and {\em vice-versa}. For
example for the lowest order {\em operators}:
\begin{eqnarray}
q \: = \: a + j\tilde{a} \: \leftrightarrow \: \left( \begin{array}{cc}
a & -\tilde{a}^{*}\\
\tilde{a} & a^{*}\end{array} \right)
\end{eqnarray}
the first column of which reproduces the symplectic state representation.
More in general for
\( {\cal H}\vert {\cal C} \; \: ( q + p\vert i; \: \: p,q \in {\cal H} ) \)
\begin{eqnarray}
q + p\vert i & = & a+j\tilde{a} + b\vert i+j\tilde{b} \vert i \nonumber \\
 & \leftrightarrow & \left( \begin{array}{cc}
a+ib & -\tilde{a}^{*}-i\tilde{b}^{*}\\
\tilde{a}+i\tilde{b} & a^{*}+ib^{*}\end{array} \right)
\end{eqnarray}
\begin{center}
(where $p=b+j\tilde{b}$) .
\end{center}
Equivalently a generic $2 \times 2$ complex matrix is given by
\begin{eqnarray}
\left( \begin{array}{cc}
a & b\\
c & d\end{array} \right) & \leftrightarrow & \frac{a+d^{*}}{2} +
j\frac{c-b^{*}}{2} \nonumber \\
 & & + \frac{a-d^{*}}{2i}\vert i + j\frac{c+b^{*}}{2i}\vert i
\end{eqnarray}
We emphasize that the above translation (eq.(\ref{b})-(\ref{e})) is limited
to {\em even complex matrices}, and that as a consequence only even complex
matrices can be translated into matrices with elements of the
form $q+p\vert i$.

We may now proceed to apply these rules. We shall first obtain the
quaternion version of a standard complex derivation of the Lorentz spinor
transformation beginning with that of the four vector $x^{\mu}$. We
shall then derive the rules for quaternion tensor products even if the
equivalent of these have already been ``guessed''\cite{del2}. We then rederive
the
above Dirac quaternion equation not from first principles but simply by
translating the standard complex equation. Finally we derive, with a
``trick'' a quaternion version of the Duffin-Kemmer-Petiau matrices which is
not only new, but for spin 0 involves {\em odd} dimensional complex
matrices, formally beyond our rules for translation.

\vspace*{11mm}

\noindent {\bf Spinor Transformations}

We briefly recall first the standard QM steps in this demonstration. Consider
the hermitian matrix X defined by
\begin{equation}
X \: = \: x^{\mu} \sigma_{\mu}
\end{equation}
where $\sigma_{0}$ is the $2 \times 2$ identity matrix and
$\sigma_{i}$ are the Pauli matrices. Then observe that $detX = x_{0}^{2} -
\vert\vec{x}\vert^{2}$ is Lorentz invariant. One can show (see ref.\cite{wkt})
that under a general Lorentz transformation
\begin{equation}
X \: \rightarrow \: X^{\prime} = LXL^{+}
\end{equation}
where
\begin{equation} \label{f}
L = exp(\frac{-i \vec{\theta} \cdot \vec{\sigma}}{2}) \hspace*{10mm} for
\: spatial \: rotations
\end{equation}
\begin{equation} \label{g}
L = exp(\frac{\vec{\xi} \cdot \vec{\sigma}}{2}) \hspace*{10mm} for \: boosts
\end{equation}
These $L$ are infact the transformation matrices for a Weyl spinor.

Now we repeat these steps with quaternions (this is more instructive than
simply translating the final result).
Define an hermitian (CSP) quaternion $X$ by
\begin{equation}
X = x_{0} + \vec{Q} \cdot \vec{x}\vert i \: ;
\hspace*{8mm} \vec{Q} \equiv (i, j, k)
\end{equation}
\begin{center}
(within the CSP $(A\vert b)^{+} = A^{+} \vert b^{+}$ when $b\in {\cal C}$) .
\end{center}
In substitution of the Lorentz invariant determinant, we first define the
``left-adjoint'' for a general $A = q + p\vert i$,
\begin{equation}
\bar{A} = q^{+} + p^{+} \vert i
\end{equation}
(n.b. that $A^{+} = q^{+} - p^{+} \vert i$ and that, as for the
adjoint, $\overline{AB} = \bar{B}\bar{A}$), then
\begin{equation}
\bar{X}X = x_{0}^{2} - \vert \vec{x}\vert^{2}
\end{equation}
As above, let us assume that under a Lorentz transformation
\[ X \: \rightarrow \: LXL^{+} \]
then
\begin{equation}
\bar{X}X \; \rightarrow \; \bar{L}^{+}\bar{X}\bar{L}LXL^{+}=\bar{X}X
\end{equation}
The necessary and sufficient condition for this to be valid, is
\begin{equation}
\bar{L}^{+}\bar{L}LL^{+} \; = \; 1 \; \: .
\end{equation}
We use the fact that both $\bar{X}X$ which is a real number and
$\bar{L}L$ which is of the form $r + s\vert i$ with $r, s$ real commute
with everything.\\
Whence, except for a (non-physical) phase factor $e^{\alpha\vert i}$ with
$\alpha$ real, the Lorentz transformations are given by
\begin{equation} \label{h}
L = exp(\frac{\vec{\theta} \cdot \vec{Q}}{2}) \hspace*{10mm} rotations
\end{equation}
or
\begin{equation} \label{i}
L = exp(\frac{\vec{\xi} \cdot \vec{Q}\vert i}{2}) \hspace{10mm} boosts
\end{equation}
or in general multiples of these transformations. Eq.(\ref{h})
and (\ref{i}) are, up to a similarity transformation, the
translations of eq.(\ref{f}) and (\ref{g}). As an aside we wish to note
that without the $q\vert i$ terms, i.e. using only simple quaternions there
is no (known) analogous derivation.

\vspace*{11mm}

\noindent {\bf The quaternion tensor products}

Consider the simplest tensor product that between two quaternion numbers
(states) $q_{1} \otimes q_{2}$. To determine its representation, we
consider its behaviour under the quaternion ``operator''
$A_{1} \otimes B_{2}$. We require that
\begin{equation}
(A_{1} \otimes B_{2})(q_{1} \otimes q_{2}) = (A_{1}q_{1}) \otimes (B_{2}q_{2})
\end{equation}
We shall use
\begin{eqnarray*}
q_{1} & \leftrightarrow & \left( \begin{array}{c} c_{1}\\
\tilde{c}_{1} \end{array} \right)
\end{eqnarray*}
and
\begin{eqnarray*}
q_{2} & \leftrightarrow & \left( \begin{array}{c} d_{2}\\
\tilde{d}_{2} \end{array} \right)
\end{eqnarray*}
Then it is immediate that
\begin{eqnarray}
q_{1} \: \otimes \: q_{2} & \leftrightarrow & \left( \begin{array}{c}
c_{1}d_{2}\\
c_{1}\tilde{d}_{2}\\ \tilde{c}_{1}d_{2}\\
\tilde{c}_{1}\tilde{d}_{2} \end{array} \right)  \nonumber \\ \\
 & \equiv & \left( \begin{array}{c} d_{2}c_{1}\\
\tilde{d}_{2}c_{1}\\ d_{2}\tilde{c}_{1} \\
\tilde{d}_{2}\tilde{c}_{1} \end{array} \right)
\end{eqnarray}
retranslated into quaternions the last matrix is identifiable with
\begin{eqnarray} \label{l}
q_{1} \: \otimes \: q_{2} & = & \left( \begin{array}{c} q_{2}c_{1}\\
q_{2}\tilde{c}_{1} \end{array} \right) \: \: .
\end{eqnarray}
Notice that the position of $q_{2}$ cannot be changed since
\[ q_{2} c_{1} \; \leftrightarrow \;  \left( \begin{array}{c} d_{2}c_{1}\\
\tilde{d}_{2}c_{1} \end{array} \right) \]
as desired, while on the contrary
\[ c_{1} q_{2} \; \leftrightarrow \;  \left( \begin{array}{c} c_{1}d_{2} \\
c_{1}^{*}\tilde{d}_{2}  \end{array} \right) \]
The above 2 component column representation is consistent with the
facts that since
\begin{eqnarray*}
A_{1} \: \otimes \: B_{2} & \leftrightarrow & ( 4 \times 4 ) \; \: complex \:
matrix \\
 & \leftrightarrow &  ( 2 \times 2 ) \; \: quaternion \: matrix
\end{eqnarray*}
it follows that,
\begin{eqnarray*}
q_{1} \: \otimes \: q_{2} & \leftrightarrow & 4 \; \: component \: complex \:
matrix \\
 & \leftrightarrow & 2 \; \: component \: quaternion \: matrix
\end{eqnarray*}
n.b. that
\begin{eqnarray} \label{w}
\left( \begin{array}{c} q_{2}c_{1}\\
q_{2}\tilde{c}_{1} \end{array} \right) \: = \: q_{2} \left( \begin{array}{c}
c_{1}\\ \tilde{c}_{1} \end{array} \right) \: \neq \: q_{2}q_{1}
\end{eqnarray}
because the right column matrix is not a complex symplectic representation
of a quaternion. Infact the matrix
\[ \left( \begin{array}{c}
c_{1}\\ \tilde{c}_{1} \end{array} \right) \]
in eq.(\ref{w}) must be considered
a {\em two component quaternion} matrix which
happens to have only complex elements. This demonstrates that one must work
either in the complex formalism or the quaternion formalism but avoid
(ambiguous) mixed formalisms. It follows that
\[ (A_{1} \otimes B_{2})(q_{1} \otimes q_{2}) =
(A_{1}q_{1}) \otimes (B_{2}q_{2}) \]
as desired. Thus if each state satisfies a generalized quaternion equation,
this will remain true for the tensor product states. It can also be shown
that the normal definition of scalar product for column matrices results in
\begin{equation}
<(q_{1} \otimes q_{2}),(q^{\prime}_{1} \otimes q^{\prime}_{2})>_{\cal C} \;
\: = \; \: <q_{1},q^{\prime}_{1}>_{\cal C}
\: <q_{2},q^{\prime}_{2}>_{\cal C}
\end{equation}
where $<f,g>_{\cal C}$ is the complex (CSP) ${\cal C}(1,i)$ projection of the
quaternion scalar product. Apart from the order of the factors $q_{1}$ and
$q_{2}$, the above definition (eq.(\ref{l})) was first given
in ref.\cite{del2}.

\vspace*{11mm}

\noindent {\bf Quaternion Dirac Equation}

This equation may be obtained by simply translating the standard Dirac
equation:
\begin{equation}
(i \gamma^{\mu} \partial_{\mu} - m ) \psi  = 0
\end{equation}
where we use the set of $4 \times 4$ gamma matrices (ref.\cite{zub})
\begin{eqnarray*}
\gamma_{0} \: = \: \left( \begin{array}{cc} {\bf 1} & 0\\
0 & {\bf -1} \end{array} \right) & \vec{\gamma} \: = \: \left(
\begin{array}{cc}
0 & \vec{\sigma}\\ -\vec{\sigma} & 0 \end{array} \right) \\
\end{eqnarray*}
\begin{center}
(with $\vec{\gamma} \equiv (\gamma_{1},\gamma_{2},\gamma_{3})$)
\end{center}
then the translation yields:
\begin{eqnarray}
\begin{array}{ccccccc}
i \gamma_{0} & \leftrightarrow & \left( \begin{array}{cc} 1\vert i & 0\\
0 & -1\vert i \end{array} \right) & \Longleftrightarrow & \gamma_{0}
& \leftrightarrow & \left( \begin{array}{cc} 1 & 0\\
0 & -1 \end{array} \right)\\ \\
i \gamma_{1} & \leftrightarrow & \left( \begin{array}{cc} 0 & -k\\
k & 0 \end{array} \right) & \Longleftrightarrow & \gamma_{1}
& \leftrightarrow & \left( \begin{array}{cc} 0 & k\vert i\\
-k\vert i & 0 \end{array} \right)\\ \\
i \gamma_{2} & \leftrightarrow & \left( \begin{array}{cc} 0 & -j\\
j & 0 \end{array} \right) & \Longleftrightarrow & \gamma_{2}
& \leftrightarrow & \left( \begin{array}{cc} 0 & j\vert i\\
-j\vert i & 0 \end{array} \right)\\ \\
i \gamma_{3} & \leftrightarrow & \left( \begin{array}{cc} 0 & i\\
-i & 0 \end{array} \right) & \Longleftrightarrow & \gamma_{3}
& \leftrightarrow & \left( \begin{array}{cc} 0 & -i\vert i\\
i\vert i  & 0 \end{array} \right)
\end{array}
\end{eqnarray}
At first sight this is not the same as the quaternion $\gamma^{\mu}$
set given in ref.\cite{rot1} (except for $\gamma^{0}$), however there exists a
similarity transformation which transforms the above set into those of
ref.\cite{rot1}:
\begin{equation}
S\gamma^{\mu}S^{-1} \; = \; \gamma^{\mu}_{ref.1}
\end{equation}
The matrix $S$ (with $S^{+} = S^{-1}$) is
\begin{eqnarray}
S \: = \: \frac{1}{\sqrt{2}} \left( \begin{array}{cc} 1+j & 0\\
0 & (1+j)\vert i \end{array} \right)
\end{eqnarray}

\vspace*{11mm}

\noindent {\bf The Duffin-Kemmer-Petiau Algebra}

The Kemmer equation is formally similar to the Dirac equation:
\begin{equation}
(i \beta^{\mu} \partial_{\mu} - m ) \psi  = 0
\end{equation}
however the $\beta^{\mu}$ are non-invertible matrices which
satisfy \cite{rom}
\begin{equation} \label{k}
\beta^{\mu} \beta^{\nu} \beta^{\lambda} + \beta^{\lambda} \beta^{\nu}
\beta^{\mu} = g^{\mu \nu} \beta^{\lambda} + g^{\lambda \nu} \beta^{\mu}
\end{equation}
Eq.(\ref{k}) guarantees that each component of $\psi$ satisfies the
Klein-Gordon equation. It can be shown that this equation describes
together a particle of spin 1 and spin 0. In the standard theory, this
equation is a false first order equation since the components of $\psi$
contain derivatives. For example the 5 components of $\psi$ (spin 0)
are identifiable with the scalar wave function $\phi$ and the four
derivatives $\partial^{\mu}\phi$. The dimension of
the $\beta$ matrices is 16 (the
algebra has 126 elements) decomposable into a trivial 1 dimensional (null)
plus 5 dimensional (spin 0) plus 10 dimensional (spin 1) representations.
The interest in finding a quaternion version to this equation is connected
to the automatic reduction of the number of
components of $\psi$. By reducing this number, it appears that there are
not enough degrees of freedom to accomodate the derivatives. One therefore
might hope to find a true ``spinor'' equation for integer spin.
This is not the case here since we are merely performing a
translation. Below we list
the spin 1 (5 dimensional) quaternion translation of the standard
$10 \times 10$ $\beta$ matrices (the points represent zeros):

\begin{eqnarray}
\begin{array}{ccc}
\beta_{1} & = & \frac{1}{2} \left( \begin{array}{ccccc}
\cdot & \cdot & \cdot & \cdot & -k-j\vert i\\
\cdot & \cdot & \cdot & \cdot & \cdot\\
\cdot & \cdot & \cdot & -i+1\vert i & -i-1\vert i\\
\cdot & \cdot & -i+1\vert i & \cdot & \cdot\\
-k+j\vert i & \cdot & -i-1\vert i & \cdot & \cdot
\end{array} \right)\\ \\
\beta_{2} & = & \frac{1}{2} \left( \begin{array}{ccccc}
\cdot & \cdot & \cdot & \cdot & -i+1\vert i\\
\cdot & \cdot & \cdot & \cdot & -k+j\vert i\\
\cdot & \cdot & \cdot & k-j\vert i & \cdot\\
\cdot & \cdot & k+j\vert i & \cdot & \cdot\\
-i+1\vert i & -k-j\vert i & \cdot & \cdot & \cdot
\end{array} \right)\\ \\
\beta_{3} & = & \frac{1}{2} \left( \begin{array}{ccccc}
\cdot & \cdot & \cdot & \cdot & \cdot\\
\cdot & \cdot & \cdot & i-1\vert i & -k-j\vert i\\
\cdot & \cdot & \cdot & i+1\vert i & \cdot\\
\cdot & i-1\vert i & i+1\vert i & \cdot & \cdot\\
\cdot & -k+j\vert i & \cdot & \cdot & \cdot
\end{array} \right)\\ \\
\beta_{0} & = & \frac{1}{2} \left( \begin{array}{ccccc}
\cdot & \cdot & \cdot & 2 & \cdot\\
\cdot & \cdot & \cdot & \cdot & 1-i\vert i\\
\cdot & \cdot & \cdot & \cdot & \cdot\\
2 & \cdot & \cdot & \cdot & \cdot\\
\cdot & 1-i\vert i & \cdot & \cdot & \cdot \end{array} \right)
\end{array}
\end{eqnarray}

For the spin 0 case since 5 is not an even number we add the trivial
solution $\beta^{\mu}=0$, which is equivalent to increasing the starting
matrices to $6 \times 6$ by adding a row and column of zeros. This
procedure can be extended to other cases in which odd dimensions are
involved, but it may not be without physical content in general. The
resulting $3 \times 3$ quaternion $\beta$ matrices are:

\begin{eqnarray}
\begin{array}{ccc}
\beta_{1} & = & \frac{1}{2} \left( \begin{array}{ccc}
\cdot & \cdot & i+1\vert i\\
\cdot & \cdot & \cdot\\
i+1\vert i & \cdot & \cdot  \end{array} \right)\\ \\
\beta_{2} & = & \frac{1}{2} \left( \begin{array}{ccc}
\cdot & \cdot & -k+j\vert i\\
\cdot & \cdot & \cdot\\
-k-j\vert i & \cdot & \cdot  \end{array} \right)\\ \\
\beta_{3} & = & \frac{1}{2} \left( \begin{array}{ccc}
\cdot & \cdot & \cdot\\
\cdot & \cdot & i+1\vert i\\
\cdot & i+1\vert i & \cdot  \end{array} \right)\\ \\
\beta_{0} & = & \frac{1}{2} \left( \begin{array}{ccc}
\cdot & \cdot & \cdot\\
\cdot & \cdot & -j-k\vert i\\
\cdot & j-k\vert i & \cdot \end{array} \right)
\end{array}
\end{eqnarray}

In conclusion we have defined a set of rules for translating from
standard QM to
a particular version of QQM. We hope that the above procedure demonstrates
the possible use of quaternions in QM, although we insist upon the non
complete nature of the translation and hence the non triviality in the
choice to adopt quaternions as the underlying number field.

\end{document}